\documentclass[
  ,draft            
  ]
  {aipproc}

\layoutstyle{6x9}

\usepackage{bm}

\begin{document}
\vspace*{-70pt}
\title[]{Azimuthal and Transverse Single Spin
Asymmetries in Hadronic Collisions
\footnote{Talk given at the ``5th International Workshop on Quantum Chromodynamics -
Theory and Experiment (QCD@Work 2010)'', 20-23 June 2010, Martina Franca - Valle d'Itria - Italy.}}

\classification{13.88.+e,~13.85.Ni,~13.60.Hb,~13.66.Bc}
\keywords      {Spin physics,~Polarization phenomena,~Inclusive processes,~Hadronic collisions}

\author{Francesco Murgia}{
  address={Istituto Nazionale di Fisica Nucleare, Sezione di Cagliari,\\
  Cittadella Universitaria di Monserrato, C.P. 170, I-09042, Monserrato (CA), Italy}
}

\begin{abstract}
We give a short overview of
the phenomenology of azimuthal and transverse single spin
asymmetries in (un)polarized high-energy hadronic collisions.
We briefly  summarize a transverse
momentum dependent, generalized parton model approach to
these polarization phenomena, and discuss some of its applications.
Finally, open points and future developments
will be outlined.
\end{abstract}
\date{}
\maketitle


\section{\label{Intro} Introduction}
It was an early common belief that transverse spin effects should play
a negligible role in high-energy hadronic reactions~\citep{Kane:1978nd}.
There are however several transverse spin effects which strongly contradict
this theoretical prejudice.
Typical examples are: a) The quark transversity distribution;
b) The large transverse polarization of hyperons produced in unpolarized
fixed-target $pN$ collisions; c) The puzzling spin-spin correlations
observed in $pp$ elastic scattering; d) The huge transverse single spin
asymmetries (SSAs) measured in the forward production of pions
in polarized $pp$ collisions; e) Several azimuthal asymmetries
measured in (un)polarized Drell-Yan (DY) processes, in semi-inclusive
deeply inelastic scattering (SIDIS) and in correlated meson-pair production in
unpolarized $e^+e^-$ collisions.

In this contribution we will concentrate on azimuthal and
transverse single spin asymmetries in hadronic collisions,
discussing recent theoretical and phenomenological
progress in the framework of the so-called
transverse momentum dependent (TMD) QCD approach,
which offers a good description and a clear understanding
of most of these phenomena.
Within this approach, a new class of leading-twist, intrinsic transverse
momentum dependent, polarized partonic distributions and fragmentation
functions are introduced which play a fundamental role in spin physics.
These TMD distributions are intimately related to
several topics of increasing interest in hadronic physics:
a) The parton orbital motion and angular momentum
inside hadrons; b) The study of hadron
structure in the impact parameter space; c) The nucleon generalized parton
distributions and deeply virtual Compton scattering; d) The light-cone
hadron wave functions.

In the sequel we will first summarize the phenomenological
motivations behind these theoretical developments.
We will then give a brief discussion of a generalized parton model approach,
with the inclusion of intrinsic parton motion and polarization effects, which has
been and is quite successful in explaining several effects
observed experimentally. We will also briefly comment on alternative
theoretical formalisms and on extensions of the TMD approach.
Finally, we will summarize recent phenomenological results of the approach,
giving some conclusions and listing open points.

This contribution is intended as an introductory mini-review
for a broad audience. Therefore, technical details will be skipped in favour
of qualitative and phenomenological aspects.
An extensive discussion on the subject and a complete list of references can be
found e.g.~in Refs.~\citep{D'Alesio:2007jt,Barone:2010}.

\section{\label{motiv} Transverse single spin asymmetries:
phenomenological motivations}
Let us first recall what a transverse single spin asymmetry is.
We will mostly consider the inclusive or semi-inclusive production
of particles with moderately large transverse momentum
in high-energy (un)polarized hadronic collisions.
Typical examples are the inclusive (single and double) meson, photon, jet production
in polarized $pp$ collisions, the polarized SIDIS and DY processes.
In all these processes there are two typical energy scales: 1) A large
scale allowing for the use of perturbative QCD techniques and factorization
schemes; 2) A small-intermediate
scale, e.g. the transverse momentum of the observed hadron
in SIDIS, of the order of few GeV at most, which keeps memory of
the intrinsic parton motion inside the hadrons involved in the process.

For strong interactions, due to parity
conservation and rotational invariance, only single spin asymmetries
with the observed spin \emph{transverse} to the production
plane survive. In this case, the transverse single spin asymmetry $A_N$,
also called left-right asymmetry, $A_{LR}$, is defined,
e.g. for the inclusive process $A^\uparrow B\to C+X$, as:
\begin{equation}
A_N(A^{\uparrow} B\to C+X) =
\frac{\,{\rm d}\sigma^{A^{\uparrow} B\to C+X}-
{\rm d}\sigma^{A^{\downarrow} B\to C+X}\,}
{\,{\rm d}\sigma^{A^{\uparrow} B\to C+X}
+{\rm d}\sigma^{A^{\downarrow} B\to C+X}\,} =
\frac{{\rm d}\Delta_N\sigma(A^{\uparrow} B\to C+X)}
{2\,{\rm d}\sigma^{\rm{unp}}(A B\to C+X)}\,,
\label{AN}
\end{equation}
where ${\rm d}\sigma(A^{\uparrow,\downarrow} B\to C+X)$ is the
transversely polarized invariant differential cross section for the process.
An analogous definition holds for the transverse polarization of hyperons (e.g.~$\Lambda$s)
inclusively produced in unpolarized hadronic collisions.

The reason why in pQCD these SSAs were expected to be negligible in inclusive
high-energy hadronic processes is that in a
standard leading-twist, collinear factorization approach, the origin of the
hadronic SSA is brought back to the SSA arising at the \emph{partonic}
level, in the hard scattering process.
Notice that by \emph{collinear factorization approach} we mean the usual
approach in which intrinsic parton motion is integrated over up to
the hard scale of the process, giving rise to the evolution
with scale of the soft functions involved, while it is neglected
in the hard perturbative scattering.

Since the transverse spin states $|\uparrow,\downarrow\rangle$ can be
written in terms of the usual $|\pm\rangle$ helicity states as
$|\uparrow,\downarrow\rangle=(1/\sqrt{2})(|+\rangle\pm i|-\rangle)$,
it is easy to see that a transverse single spin asymmetry is
related to the imaginary part of the interference term between
off-diagonal helicity amplitudes:
\begin{equation}
A_N \propto |\langle|\dots|\uparrow\rangle|^2 -
  |\langle|\dots|\downarrow\rangle|^2 \propto
  {\rm Im}\,\langle|\dots|\pm\rangle\langle|\dots|\mp\rangle^*\,.
\label{im-int}
\end{equation}

Since at tree level helicity amplitudes
are real (up to an overall phase), and the pQCD
massless $qg$ coupling preserves helicity,
it is easy to see that the partonic
SSA, $\hat a_N$, is strongly suppressed for large energy and
transverse momentum scales~\citep{Kane:1978nd}:
\begin{equation}
\hat{a}_N = \frac{{\rm d}\hat\sigma^{\,a^\uparrow b\to cd}
-{\rm d}\hat\sigma^{\,a^\downarrow b\to cd}}
{{\rm d}\hat\sigma^{\,a^\uparrow b\to cd} +
{\rm d}\hat\sigma^{\,a^\downarrow b\to cd}}
\propto \alpha_s(\hat s)\,\frac{m_q}{\hat{s}}
\sim \alpha_s(p_{T})\,\frac{m_q}{p_T}\,.
\label{an-parton}
\end{equation}

Against this common wisdom, a huge amount of extremely puzzling results
on transverse hyperon polarization in fixed-target unpolarized
proton-nucleus collisions were collected during the 70s~\citep{Heller:1996pg}.
Since these data were at relatively low c.m.~energy and transverse momentum,
they were mainly interpreted as soft nonperturbative effects.
Starting from the 90s, however, the E704
Collaboration at Fermilab measured huge SSAs for the
process $p^\uparrow p\to \pi+ X$ at $\sqrt{s}\sim 20 $ GeV and
$0.7 < p_T < 2.0$ GeV in the forward region
($x_F = 2 p_L/\sqrt{s} > 0.4$)~\citep{Adams:1991cs}.
These results have been recently confirmed
by the STAR Collaboration at RHIC~\citep{Abelev:2008qb}
at much larger energies ($\sqrt{s}=200$ GeV),
again in the forward region (the pion SSA is almost negligible in the
central and negative (pseudo)rapidity regions)
and for $p_T$ up to $\sim 4$ GeV.
While leading-twist NLO collinear pQCD gives a fair account of the
corresponding unpolarized cross sections in the same RHIC kinematical
regime, it is unable to explain these huge SSAs.
At that time, the results of the E704 collaboration triggered renewed theoretical
and experimental efforts aiming at an understanding of these phenomena
and of the physical mechanisms behind them.

Apart from the already mentioned RHIC extensive research program on
spin physics~\citep{Bunce:2000uv}, several azimuthal and single spin asymmetries
have been studied and measured in polarized SIDIS processes
by the HERMES-DESY~\citep{Airapetian:2009ti} and
COMPASS-CERN~\citep{Alekseev:2010rw} collaborations,
in the Drell-Yan process~\citep{Zhu:2006gx}
and in almost back-to-back
two-particle correlations in $e^+e^-$ collisions at Belle~\citep{Seidl:2008xc}.
In several cases these asymmetries result to be sizable
and difficult to explain in the usual collinear approach.
\section{\label{theo} The TMD generalized parton model approach}
{}From the theoretical point of view, two different approaches
have been proposed:\\
1) The so-called twist-three collinear approach~\citep{Qiu:1991pp} works along
the lines of the collinear pQCD factorization methods with the
necessary inclusion of higher-twist quark-gluon correlation
functions and a new class of twist-three
parton distribution and fragmentation functions.
While this method is less problematic from the point of
view of the validity of the factorization procedure
(in particular for single inclusive particle
production in hadronic collisions) it has presently the problem
that unpolarized cross sections (entering
the denominator of the SSAs) can only be evaluated at leading twist level.\\
2) The second approach, which will be discussed at length in this
contribution, is the so-called transverse momentum dependent
QCD approach. In this approach, the intrinsic parton motion of
partons inside initial hadrons and of produced hadrons w.r.t.~the
fragmenting final partons is not integrated over and is
taken into account explicitly. Intrinsic parton motion plays
a fundamental role in allowing for sizable azimuthal and
SSAs as those discussed above: e.g., it is the possible azimuthal
asymmetry in the distribution of unpolarized partons around the direction of
motion of the parent, transversely polarized, proton which
can explain the huge pion SSAs observed at Fermilab and RHIC
in the moderately large $p_T$ region. This mechanism, called
the Sivers effect, was first suggested by D.~Sivers~\citep{Sivers:1989cc}.

The first phenomenological realization of the TMD approach,
that we will call generalized parton model, takes into account
intrinsic parton motion both in the soft and hard components
of the factorized cross section, including polarization effects
and adopting the helicity formalism~\citep{Anselmino:2005sh}.
Factorization is assumed as a reasonable starting point.
Later developments of the TMD approach~\citep{Boer:2003cm}
led to the so-called
TMD color gauge invariant approach, where appropriate
gauge links (Wilson lines), preserving gauge invariance, are
introduced in the hadronic correlators:
the corresponding (perturbative) gluon exchanges among
partons before and after the hard scattering and the
hadron remnants give rise to the
imaginary interference terms required for a non vanishing SSA.

Since in this contribution we are mainly interested in the
basic ideas of the formalism, we will limit our discussion
to the generalized parton model approach, which is the
most intuitive and easy to illustrate.
In this approach, e.g.~the invariant differential
cross section for the doubly polarized single
particle inclusive production in hadronic collisions,
$A(S_A)\,B(S_B)\to C+X$,
can be written as~\citep{Anselmino:2005sh}:
\begin{eqnarray}
\frac{E_C \, {\rm d}\sigma^{(A,S_A) + (B,S_B) \to C + X}}
{{\rm d}^{3} \bm{p}_C}
 &= & \sum_{a,b,c,d, \{\lambda\}}
 \! \! \int \frac{{\rm d}x_a \,
{\rm d}x_b \, {\rm d}z}{16 \pi^2 x_a x_b z^2  s} \;
{\rm d}^2 \bm{k}_{\perp a} \, {\rm d}^2 \bm{k}_{\perp b}\,
 {\rm d}^3 \bm{k}_{\perp C}\,
\delta(\bm{k}_{\perp C} \cdot \hat{\bm{p}}_c) \, \nonumber \\
 \!\!& \times & \! \! J(\bm{k}_{\perp C})
 \rho_{\lambda^{\,}_a,
\lambda^{\prime}_a}^{a/A,S_A} \, \hat f_{a/A,S_A}(x_a,\bm{k}_{\perp a})
\> \rho_{\lambda^{\,}_b, \lambda^{\prime}_b}^{b/B,S_B} \,
\hat f_{b/B,S_B}(x_b,\bm{k}_{\perp b}) \nonumber\\
\! \! &\times& \! \! \hat M_{\lambda^{\,}_c, \lambda^{\,}_d;
 \lambda^{\,}_a, \lambda^{\,}_b} \,
\hat M^*_{\lambda^{\prime}_c, \lambda^{\,}_d; \lambda^{\prime}_a,
\lambda^{\prime}_b} \> \delta(\hat s + \hat t + \hat u) \> \hat
D^{\lambda^{\,}_C,\lambda^{\,}_C}_{\lambda^{\,}_c,
\lambda^{\prime}_c}(z,\bm{k}_{\perp C})\,,
\label{gen1}
\end{eqnarray}
where: $x_{a,b}$, $z$, and $\bm{k}_{\perp a,b,C}$ are respectively
the light-cone momentum fractions and the intrinsic transverse momenta
of the initial partons $a$, $b$, inside hadrons $A$, $B$, and of the final
observed hadron $C$ inside the fragmentation jet of the scattered
parton $c$; $J(\bm{k}_{\perp C})$ is a kinematical factor;
$\rho_{\lambda^{\,}_a,\lambda^{\prime}_a}^{a/A,S_A}$
is the helicity density matrix of parton $a$ inside hadron $A$;
the quantity $\rho_{\lambda^{\,}_a,\lambda^{\prime}_a}^{a/A,S_A}
\hat f_{a/A,S_A}(x_a,\bm{k}_{\perp a})$ encodes complete
information on the polarization state of parton $a$ and
is related to the leading-twist TMD parton distribution
functions which generalize the usual collinear PDFs;
analogously for parton $b$ inside hadron $B$;
the $\hat M_{\lambda^{\,}_c, \lambda^{\,}_d;
 \lambda^{\,}_a, \lambda^{\,}_b}$'s are the LO helicity
 scattering amplitudes for the hard partonic process
 $ab\to cd$; finally, $D^{\lambda^{\,}_C,\lambda^{\,}_C}_{\lambda^{\,}_c,
\lambda^{\prime}_c}(z,\bm{k}_{\perp C})$ is the soft
function describing the fragmentation process of the polarized parton $c$
into the observed hadron $C$. In the sequel, for simplicity
we will only consider the case of spinless
or unpolarized final particles, for which this soft
function simplifies to $D^{C}_{\lambda^{\,}_c,
\lambda^{\prime}_c}(z,\bm{k}_{\perp C})$.

The polarization state of the initial parton $a$ (and analogously
for parton $b$) depends on the polarization state of the parent
hadron $A$, which is fixed by the experimental conditions (we will
have in mind spin-1/2 initial hadrons in the sequel) and on
the soft (polarized) process $A(S_A) \to a(s_a) + X$:
\begin{equation}
\rho_{\lambda^{\,}_a,\lambda^{\prime}_a}^{a/A,S_A}
\hat f_{a/A,S_A}(x_a,\bm{k}_{\perp a}) =
\sum_{\lambda^{\,}_A,\lambda^{\prime}_A}\,
\rho_{\lambda^{\,}_A,\lambda^{\prime}_A}^{A,S_A}
\,\hat{F}^{\lambda^{\,}_a,\lambda^{\prime}_a}
_{\lambda^{\,}_A,\lambda^{\prime}_A}
(x_a,\bm{k}_{\perp a})\,.
\label{F}
\end{equation}
\indent The transverse momentum dependent soft functions
$\hat{F}^{\lambda^{\,}_a,\lambda^{\prime}_a}
_{\lambda^{\,}_A,\lambda^{\prime}_A}$ are related
to the leading twist hadronic correlator and the
typical hand-bag diagrams for DIS. In principle
there are 16 different functions.
Rotational invariance and parity conservation for
strong interactions reduce this number to eight independent
TMD distribution functions (to be compared
with the three fundamental parton distributions
in the collinear approach).
Bearing in mind that upper(lower) helicity indexes refer to
parton(hadron) respectively, and that (off-)diagonal helicity
combinations refer to (transversely)longitudinally polarized
particles, these functions can be combined in a way
which clarify their physical meaning;
e.g.~for quark partons:
1) $\hat{F}^{++}_{++} \pm \hat{F}^{--}_{++}$
are real quantities related to the unpolarized
(longitudinally polarized) quark distributions;
2) $\hat{F}^{+-}_{+-} \pm \hat{F}^{-+}_{+-}$
are also purely real quantities and are related
to the quark transversity distribution;
3) $\hat{F}^{++}_{+-} \pm \hat{F}^{--}_{+-}$
describe respectively an unpolarized (longitudinally polarized)
quark inside a transversely polarized hadron and are associated to
the Sivers function mentioned above and to the $g_{1T}^\perp$
distribution;
4) $\hat{F}^{+-}_{++} \pm \hat{F}^{+-}_{--}$
describe a transversely polarized quark
inside an unpolarized(longitudinally polarized)
hadron and are known respectively as the Boer-Mulders (BM)
function~\citep{Boer:1997nt} and the $h_{1L}^\perp$ distribution.

Notice that in the collinear, $\bm{k}_\perp$-integrated
configuration, the only surviving functions are
$\hat{F}^{++}_{++} \pm \hat{F}^{--}_{++}$ and  $\hat{F}^{+-}_{+-}$,
that is the three fundamental quark parton distributions,
respectively the unpolarized, longitudinally and transversely
polarized distributions.
All other functions, due to the presence of a transverse polarization
(either of the quark or of the hadron or both)
w.r.t.~the plane containing the quark and the parent
hadron, can be azimuthally asymmetric around the direction of
motion of the hadron.
It is this azimuthal asymmetry, at the partonic level, that can
give rise, in processes where a relatively small transverse
momentum scale is measured, to a correlation among intrinsic
motion and polarization effects which can survive at the
hadronic level even at leading twist, as for example in SIDIS
and Drell-Yan processes.
For inclusive single particle production in $pp$ collisions
the situation is slightly more involved: in order to have
a non vanishing asymmetry one needs to keep into account
intrinsic parton motion also in the hard processes.
This can cast some doubts on the validity
of the factorization procedure and effectively makes
the asymmetry \emph{at hadronic level} a twist-three effect.

Analogous arguments can be used for the leading twist
TMD parton fragmentation functions. In this case one
finds that, for spinless or unpolarized particles,
only two TMD functions survive: one related to the
usual collinear unpolarized FF, and a second one,
the Collins fragmentation
function~\citep{Collins:1992kk}, which describes
the azimuthal asymmetry in the distribution of
hadrons (inside the fragmentation jet)
around the direction of motion of the fragmenting parton.
For spin-1/2 particles, e.g.~hyperons, in close
analogy with the distribution sector, there are
instead eight independent leading twist, TMD
fragmentation functions.
\section{\label{pheno} Phenomenology}
Let us now summarize and briefly comment on the
TMD functions most relevant from the phenomenological
point of view. Essentially there are two of them in
the distribution sector and two in the fragmentation
sector:\\
\noindent 1) The chiral-even, naively T-odd,
Sivers distribution function~\citep{Sivers:1989cc}:
\begin{equation}
\Delta \hat{f}_{q/p^\uparrow}(x,\bm{k}_\perp)=
\hat{f}_{q/p^\uparrow}(x,\bm{k}_\perp)-
\hat{f}_{q/p^\downarrow}(x,\bm{k}_\perp)=
\hat{f}_{q/p^\uparrow}(x,\bm{k}_\perp)-
\hat{f}_{q/p^\uparrow}(x,-\bm{k}_\perp)\,.
\label{siv}
\end{equation}
The Sivers function describes the azimuthal asymmetry in the
distribution of unpolarized  quarks around the direction of
motion of the transversely polarized parent proton. It
plays a relevant role for SSAs
in polarized $AB\to C+ X$, SIDIS and DY processes.\\
\noindent 2) The chiral-odd, naively T-odd,
Boer-Mulders distribution~\citep{Boer:1997nt}:
\begin{equation}
\Delta \hat{f}_{q^\uparrow/p}(x,\bm{k}_\perp)=
\hat{f}_{q^\uparrow/p}(x,\bm{k}_\perp)-
\hat{f}_{q^\downarrow/p}(x,\bm{k}_\perp)=
\hat{f}_{q^\uparrow/p}(x,\bm{k}_\perp)-
\hat{f}_{q^\uparrow/p}(x,-\bm{k}_\perp)\,.
\label{boer}
\end{equation}
It describes the azimuthal asymmetry in the
distribution of transversely polarized quarks
around the direction of
motion of the unpolarized parent proton, and
plays a role for several azimuthal asymmetries
in unpolarized $AB\to C+ X$, SIDIS and DY processes.\\
\noindent 3) The chiral-odd, naively T-odd,
Collins fragmentation function~\citep{Collins:1992kk}:
\begin{equation}
\Delta \hat{D}_{h/q^\uparrow}(z,\bm{k}_\perp)=
\hat{D}_{h/q^\uparrow}(z,\bm{k}_\perp)-
\hat{D}_{h/q^\downarrow}(z,\bm{k}_\perp)=
\hat{D}_{h/q^\uparrow}(z,\bm{k}_\perp)-
\hat{D}_{h/q^\uparrow}(z,-\bm{k}_\perp)\,,
\label{collins}
\end{equation}
which is related to the
azimuthal asymmetry in the distribution of unpolarized
hadrons around the direction of
motion of the transversely polarized fragmenting quark.
It plays a major role for azimuthal and spin asymmetries
in (un)polarized $AB\to h+ X$, SIDIS, DY, and
$e^+e^-\to h_1 h_2+ X$ processes.\\
\noindent 4) The chiral-even, naively T-odd,
``Polarizing'' fragmentation function~\citep{Anselmino:2000vs}:
\begin{equation}
\Delta \hat{D}_{h^\uparrow/q}(z,\bm{k}_\perp)=
\hat{D}_{h^\uparrow/q}(z,\bm{k}_\perp)-
\hat{D}_{h^\downarrow/q}(z,\bm{k}_\perp)=
\hat{D}_{h^\uparrow/q}(z,\bm{k}_\perp)-
\hat{D}_{h^\uparrow/q}(z,-\bm{k}_\perp)\,,
\label{polar}
\end{equation}
describing the
azimuthal asymmetry in the distribution of transversely
polarized, spin-1/2 hadrons $h$ (e.g.~$\Lambda$ hyperons),
around the direction of motion of an unpolarized fragmenting quark.
It plays a relevant role for the transverse hyperon polarization
in unpolarized $AB\to h+ X$ and SIDIS processes.

Over the last years the TMD generalized parton model has been
extensively used in phenomenological analyses of a large
set of measured azimuthal and single spin asymmetries,
including data on $A_N(p^\uparrow p\to\pi+ X)$ and
from polarized SIDIS processes for the Sivers
and Collins asymmetries, and from unpolarized DY and
$e^+e^-\to h_1h_2+X$ processes, involving respectively,
among others, the Boer-Mulders distribution and the Collins effect.
In SIDIS processes, $\ell p \to \ell^\prime h +X$,
one looks at the inclusive production, in the
virtual photon - target proton c.m.~reference frame,
of hadrons with transverse momentum $P_{hT}$ of the
order of 1 GeV, and measures the differential cross section
as a function of $P_{hT}$ and of the azimuthal
angles of the hadron transverse momentum and transverse
spin, measured w.r.t. the leptonic plane.
We define the azimuthal asymmetries:
\begin{equation}
A_{S_B\,S_T}^{W(\phi_h,\phi_S)} =
2\langle\,W(\phi_h,\phi_S)\,\rangle =
2\,\frac{\int {\rm d}\phi_h {\rm d}\phi_S\, W(\phi_h,\phi_S)\,
\,[\,{\rm d}\sigma(\phi_h,\phi_S)-{\rm d}\sigma(\phi_h,\phi_S+\pi)\,]}
{\int {\rm d}\phi_h {\rm d}\phi_S\,[\,{\rm d}\sigma(\phi_h,\phi_S)
+{\rm d}\sigma(\phi_h,\phi_S+\pi)\,]}\,,
\label{azim-asy}
\end{equation}
where $S_B$, $S_T$ are respectively the beam ($S_B= U, L$) and
target ($S_T=U,L,T$) polarizations and $W(\phi_h,\phi_S)$
is an appropriate circular function,
e.g.~$W=\sin(\phi_h\mp \phi_S)$ respectively for the Sivers
and the Collins effects~\citep{Airapetian:2009ti,Alekseev:2010rw}.
Analogously, in $e^+e^-\to h_1 h_2+ X$ processes,
one looks at azimuthal correlations for two almost back-to-back
hadrons (mainly pions) produced in the two-jet fragmentation of the
high-energy parent $q\bar q$ pair.
Again, in the $e^+e^-$ c.m.~frame, one can measure
the asymmetry $\propto \langle \cos(\phi_1+\phi_2)\rangle$,
where $\phi_1$ and $\phi_2$ are the azimuthal
angles of the two hadron momenta w.r.t.~the plane containing the
lepton beams and the jet thrust axis~\citep{Seidl:2008xc}. This asymmetry
is related to the Collins effect in the fragmentation process.

Using combined data from SIDIS and $e^+e^-$ processes,
an updated set of parameterizations for the TMD
Sivers~\citep{Anselmino:2008sga} and (for the first time)
transversity distributions and for the
Collins function~\citep{Anselmino:2008jk} has been
extracted and made available for estimates of asymmetries in different
processes and kinematical configurations accessible presently or in the near
future by a number of experimental setups.

The case of single inclusive particle production in polarized
$pp$ collisions, which historically triggered the theoretical
and phenomenological activity on transverse SSAs,
is in fact much more involved:
as we said before, the SSA is in this case a twist-three
effect in a $1/p_T$ expansion, and several mechanisms, in particular
the Sivers and Collins effects, can be present on the same basis.
Moreover, for the distributions, the region of light-cone momentum fraction,
$x$, covered by SIDIS data on azimuthal asymmetries
is relatively low ($x\leq 0.3$).
As a consequence, all parameterizations available for TMD distributions,
in particular for the Sivers and transversity distributions, are
plagued by large uncertainties in the large $x$ region, which is the
region relevant for the huge forward SSAs observed in $pp$ collisions.
From this point of view, the study of azimuthal asymmetries in the
distribution of leading pions inside a jet in
$p^\uparrow p\to {\rm jet}+\pi+X$  processes can be quite useful
since it allows, analogously to SIDIS processes, to disentangle
among the various contributions. Work in this direction has been
already done~\citep{Yuan:2007nd}
and further extensions are currently in progress~\citep{dmp:2010}.
\begin{table}
\begin{tabular}{lccccccc}
\hline
\tablehead{1}{c}{b}{Process}
  & \tablehead{1}{c}{b}{Twist}
  & \tablehead{1}{c}{b}{Sivers}
  & \tablehead{1}{c}{b}{Collins}
  & \tablehead{1}{c}{b}{B-M}
  & \tablehead{1}{c}{b}{Pol. FF}
  & \tablehead{1}{c}{b}{Theor.\\Status}
   & \tablehead{1}{c}{b}{Discr.\\Power}  \\
\hline
SIDIS $(\ell p\to \ell'\,h+X)$
                        & 2 & $\bullet$ & $\bullet$ & $\bullet$ & $\bullet$ & **** & **** \\
Drell-Yan $(AB\to\ell^-\ell^++X)$
                        & 2 & $\bullet$ &           & $\bullet$ &           & **** & **** \\
$e^+e^-\to h_1\,h_2+ X$ & 2 &           & $\bullet$ &           & $\bullet$ & **** & **** \\
$AB\to h + X$           & 3 & $\bullet$ & $\bullet$ & $\bullet$ & $\bullet$ & **   & **   \\
$AB\to\gamma + X$       & 3 & $\bullet$ &           & $\bullet$ &           & **   & ***  \\
$AB\to h_1\,h_2 + X$    & 2 & $\bullet$ & $\bullet$ & $\bullet$ & $\bullet$ & ***  & ***  \\
$AB\to{\rm jet}  + X$   & 3 & $\bullet$ &           & $\bullet$ &           & **   & **** \\
$AB\to{\rm jet}\,\,h+ X$& 2 & $\bullet$ & $\bullet$ & $\bullet$ & $\bullet$ & ***  & **** \\
$AB\to{\rm jet}\,\,\gamma+ X$
                        & 2 & $\bullet$ &           & $\bullet$ &           & ***  & **** \\
\hline
\end{tabular}
\caption{A summary of the most interesting and
experimentally accessible processes; for each of them,
the TMD effects involved, the theoretical status and
the usefulness in discriminating among different mechanisms
are indicated.}
\label{tab:a}
\end{table}

In Table~\ref{tab:a} we summarize some of the
most phenomenologically interesting processes,
specifying the relevant mechanisms involved, their theoretical status
concerning e.g.~factorization, and their discriminating power,
that is their usefulness in disentangling among
the various mechanisms.
Far from being a complete summary,
this table gives however an idea of the
present phenomenological activity in the field
(see also Refs.~\citep{D'Alesio:2007jt,Barone:2010}).
\section{\label{outlook} Outlook and open points}
Let us finally conclude by simply quoting some of the most
relevant open points in this field and of the most promising
developing topics: 1) For inclusive single and (partially) double
inclusive particle production in hadronic collisions factorization
remains to be proved and poses several difficulties;
2) Properties of evolution with scale of the TMD distribution
and fragmentation functions are not well understood and much remains
to be done on these aspects; 3) A consistent inclusion of all unknown
soft factors (hopefully spin independent) from soft-gluon
radiation has to be performed yet;
4) Related to this, the potential suppression of
azimuthal asymmetries coming from Sudakov factors
has been only partially investigated;
5) The role of parton offshellness and fully
unintegrated parton correlation functions needs
further study; 6) From a more phenomenological side,
experimental tests of the universality (breaking?) of the
TMD distributions, by carefully comparing different processes,
are crucial at the present stage.

Hopefully in the
near future more refined theoretical results, improved
parameterizations and phenomenological constraints
will help us in clarifying several points raised above,
improving our understanding of the transverse structure
and of parton orbital motion inside hadrons.

\end{document}